\documentclass{ptapap}

\author{Henryka Netzel}[UW]
\author{Rados\l{}aw Smolec}[CAMK]
\affil[UW]{Warsaw University Astronomical Observatory\\
Al. Ujazdowskie 4, 00--478 Warszawa, Poland}
\affil[CAMK]{Nicolaus Copernicus Astronomical Center\\
ul. Bartycka 18, 00--716 Warszawa, Poland}

\title{Non-radial Pulsations in RR~Lyrae Stars from the OGLE Collection}

\begin{document}

\maketitle

\begin{abstract}
RR~Lyrae stars are classical pulsating stars. They pulsate mostly in the radial fundamental mode (RRab stars), in the radial first overtone mode (RRc stars), or in both modes simultaneously (RRd stars). Collection of variable stars from the Optical Gravitational Lensing Experiment (OGLE) contains more than $38\,000$ RR~Lyrae stars from the Galactic bulge. We analysed these data for RRc and RRd stars. We have found new members of radial--non-radial double-mode RR~Lyrae stars, with characteristic period ratio of the two modes around $0.61$. We increased the number of known RR~Lyrae stars of this type by a factor of 8. 

We have also discovered another group of double-mode RR~Lyrae stars. They pulsate in the first overtone and in another, unidentified mode, which has period longer than period of the undetected fundamental mode. The period ratios tightly cluster around $0.686$. These proceedings are focused on this puzzling group. In particular, we report eight new members of the group.

\end{abstract}

\section{Introduction}
\begin{figure}[!ht]
\centering
\includegraphics[width=0.8\textwidth]{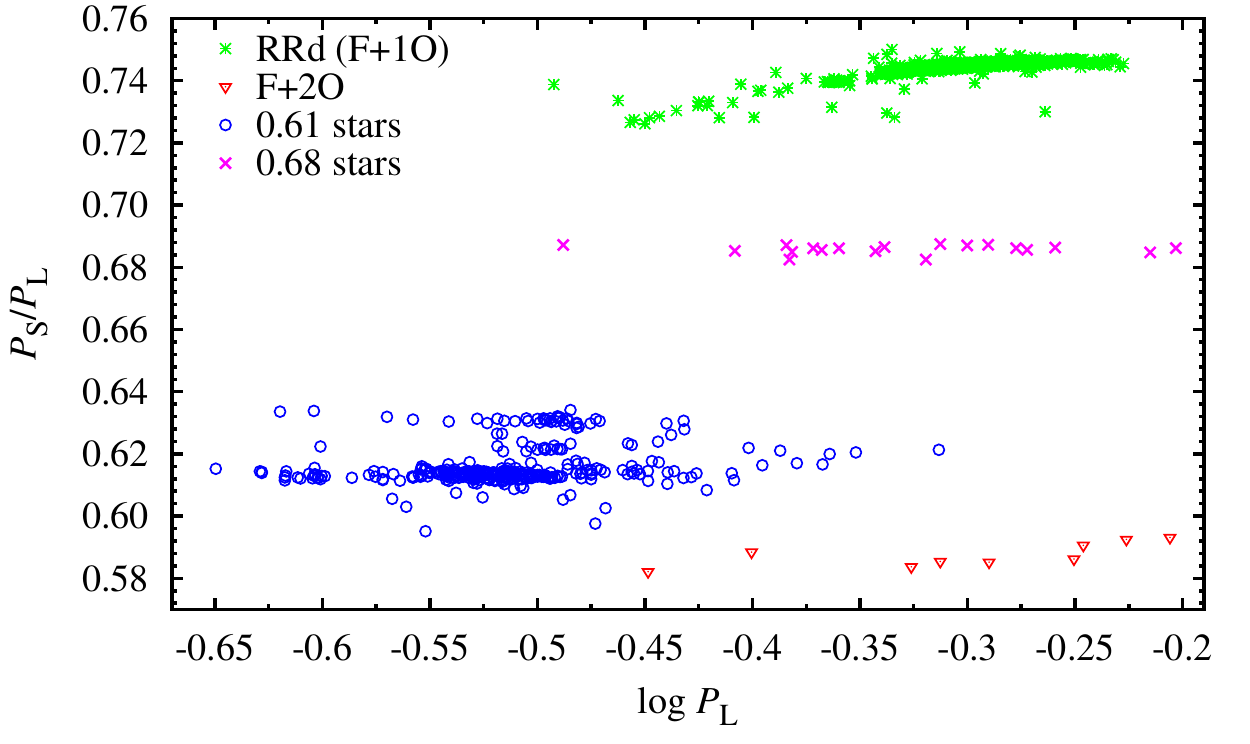}
\caption{Multi-mode pulsations of RR~Lyrae stars in the Petersen diagram.} 
\label{fig.pet} 
\end{figure}

RR~Lyrae stars pulsate mostly in the radial fundamental mode or in the radial first overtone mode. Double-mode radial pulsators are also known. They pulsate simultaneously either in the fundamental mode and in the first overtone (green asterisks in Fig.~\ref{fig.pet}), or in the fundamental mode and in the second overtone (red triangles in Fig.~\ref{fig.pet}). Observations revealed another group of double-mode RR~Lyrae stars. These stars pulsate in the first overtone (RRc or RRd) and have another signal of higher frequency (shorter period). Period ratio of the additional mode to the first overtone is around $0.61$. The additional variability cannot correspond to any radial mode \citep{pamsm15}. Until recently, this group was not numerous: 23 stars were discovered in various stellar systems both in the ground and in the space observations \citep[for a summary see][]{pamsm15}. Additional 18 stars of this type were discovered in M3 by \cite{jurcsik_M3}.

We have analysed photometry for RRc stars from the Galactic bulge collected by the OGLE project \citep{ogle-iv}. We have significantly increased the number of double-mode stars with $P_{\rm x}/P_{\rm 1O}\approx0.61$ \citep[][]{061oiii,061oiv}. Thanks to our analysis, three separate sequences formed by these stars in the Petersen diagram  were revealed (blue asterisks in Fig.~\ref{fig.pet}). Explanation for this puzzling group of double-periodic pulsators was proposed recently by \cite{dziembowski}.

These proceedings are focused on another exciting discovery we have made in the OGLE data \citep{068}. We have detected additional, unexpected signals in a group of RRc stars (magenta crosses in Fig.~\ref{fig.pet}). Additional signal has frequency lower than the frequency of the first overtone; even lower than the frequency of the fundamental mode, which is undetected in these stars. Explanation of the nature of these signals faces difficulties. For short discussion see \cite{068} and \cite{dziembowski}.

\section{Data analysis}

Our analysis of RR~Lyrae stars was focused on a search of additional low-amplitude signals. In the OGLE collection of variable stars there are more than $10\,000$ RRc stars \citep{ogle-iv-bulge}. First, we decided to focus on stars which were most frequently observed. These stars are located in the OGLE fields 501 and 505 \citep[see position of observational fields in fig.~15 in][]{ogle-iv}. Selected stars have more than $8\,000$ data points from four observational seasons of OGLE-IV. There are $485$ RRc stars in the sample. They were analysed manually, using standard consecutive prewhitening technique, with the dedicated software written by the authors. For some stars, after prewhitening with the first overtone and its harmonics, residual signal at the frequency of the first overtone remained. It may correspond to long-term period changes in these stars or to long-term amplitude and/or phase modulation (the Blazhko effect). For these stars, to increase the frequency resolution, we combined OGLE-IV data with data available from the previous phases of the OGLE project \citep{ogle-iii-bulge}. Increased data length allowed us to resolve close frequencies and to detect long-period modulation in a few stars.

Analysis of the whole sample of more than $10\,000$ RRc stars in the Galactic bulge is ongoing. Some of preliminary results are reported in these proceedings.

\section{New group of double-mode stars}
 
During the analysis of the selected sample of 485 RRc stars, we detected an additional signal in 11 stars (2 per cent of the analysed sample). Frequency of the additional signal is lower than the frequency of the fundamental mode, which is not detected in these stars. In the Petersen diagram in Fig.~\ref{fig.pet}, stars of the new group are marked  with magenta crosses. They form a horizontal sequence and tightly cluster around $0.686$. Typical amplitude of the additional signal corresponds to few per cent of the first overtone amplitude (and does not exceed 10 per cent). 

A literature search revealed one additional RRc star of this type, KIC945311, which was observed with {\it Kepler} space telescope and analysed by \cite{pamsm15}. In this star additional 0.61 mode was discovered. For one of the additional low-amplitude periodicities \citep[designated as $f_5$ in tab. 7 in ][]{pamsm15} we have found $P_{\rm 1O}/P_{5}=0.6867$. Thus, this star fits the progression in the Petersen diagram very well and is also a member of the new group. These results were described in detail in \cite{068}.

Preliminary analysis of the full sample of $10\,826$ OGLE RRc stars stars revealed eight new stars that belong to this group. They are reported here for the first time. Altogether there are 19 members of the new group discovered in the OGLE data and one star discovered in the {\it Kepler} data. Basic properties of all stars are collected in Tab.~\ref{tab.sh}. First part of the table contains stars described in \cite{068}, including one star observed with {\it Kepler} telescope. Second part of the table contains data on 8 newly discovered stars. We note that in all these stars we detect signals at combination frequency with the first overtone, which proves that additional signal is intrinsic to these stars and is not due to contamination.

\begin{table}[t!h!]
\centering
\caption{RRc stars with additional long-period signal. Subsequent columns contain: periods of the first overtone and of the additional signal, their ratio and amplitude ratio. Last column contains remarks. First part of the table lists stars described in detail in \cite{068}. Last part of the table contains recent discoveries in the OGLE data, reported here for the first time.}
\smallskip
\label{tab.sh}
\begin{tabular}{lllllll}
\hline 
  star & $P_{\rm 1O}$ (d) & $P_{\rm x}$ (d) &$P_{\rm 1O}/P_{\rm x}$&  $A_{\rm x}/A_{\rm 1O}$ &  remarks\\ 
\hline
04994 & 0.3622954(3) & 0.52797(1) & 0.68619 &  0.048 &  a \\
05080 & 0.2996982(1) & 0.436810(5) & 0.68611 &  0.048 & e  \\
06970 & 0.42988998(7) & 0.626497(9) & 0.68618 & 0.012 & a, e \\
07127 & 0.3778854(3) & 0.55057(1) & 0.68635 &  0.050 &   a \\
07653 & 0.31118879(6) & 0.454151(6) & 0.68521 &  0.025 &   a \\
08748 & 0.29153824(7) & 0.424892(3) & 0.68615 &  0.039 &   a, b \\
09146 & 0.35215799(3) & 0.5124233(5) & 0.68724 &  0.028 &  c, e \\
09217 & 0.29391346(2) & 0.428719(1) & 0.68556 &  0.026 &   a, e \\
09426 & 0.22339120(4) & 0.325069(6) & 0.68721 &  0.016 &  a, e \\
10100 & 0.4173796(2) & 0.609497(8) & 0.68479 & 0.060 &  a, e \\
32196 & 0.2677486(1) & 0.390699(7) & 0.68531 & 0.039 &  a \\
 KIC 9453114 & 0.3660809 & 0.5330831 & 0.68672 &  0.004  & e \\ 
\hline 
01064 & 0.3148614(1) & 0.458621(7) & 0.68654 & 0.030 & e\\
07673 & 0.3662894(1) & 0.534248(4) & 0.68562 & 0.060 & a, e\\
08076 & 0.28448202(6) & 0.415333(1) & 0.68495 & 0.100 & e\\
09671 & 0.28267093(4) & 0.414171(4) & 0.68250 & 0.020 & a, e\\
10427 & 0.3442403(1) & 0.501076(4) & 0.68700 & 0.057 & a, e\\
14775 & 0.3346791(1) & 0.48682(1) & 0.68748 & 0.025 & e\\
30601 & 0.2836980(4) & 0.412854(9) & 0.68716 & 0.077 & e, a, d\\
36675 & 0.3271380(1) & 0.479323(6) & 0.68250 & 0.048 & e\\
\hline
\multicolumn{6}{l}{a -- additional signal close to $f_{\rm 1O}$; b -- harmonic of $f_{\rm x}$;}\\
\multicolumn{6}{l}{c -- additional signal close to $f_{\rm x}$; d -- additional signal in power spectrum}\\
\multicolumn{6}{l}{e -- combination frequency between the first overtone and additional signal}\\
\hline
\end{tabular}
\end{table}
 
\begin{figure}[!ht]
\centering
\includegraphics[width=0.8\textwidth]{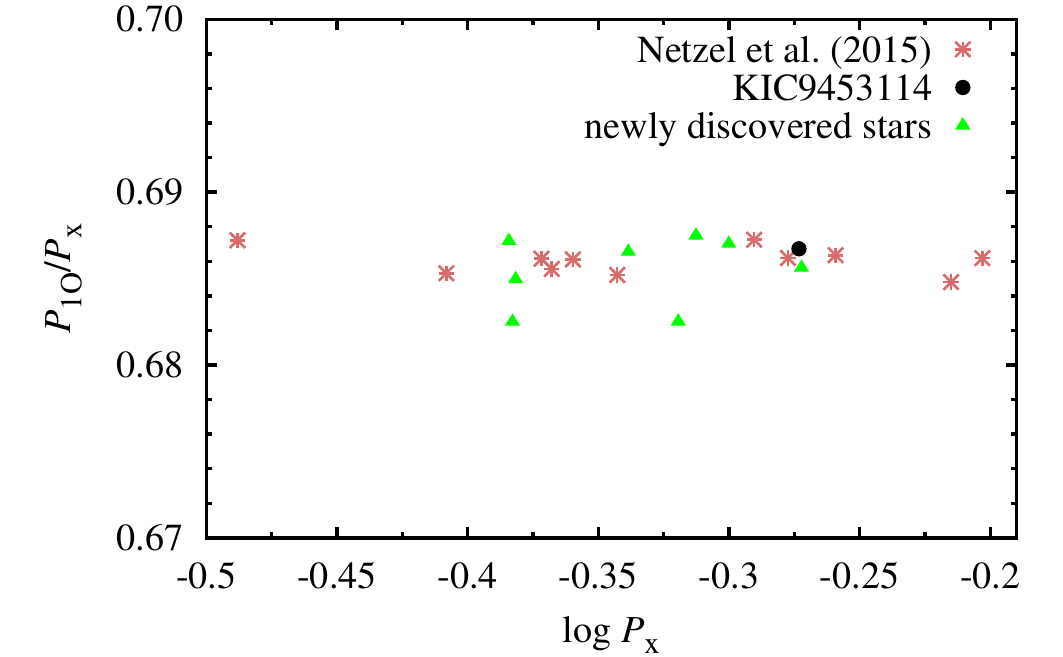}
\caption{Petersen diagram for RRc stars with additional, long-period signal. New detections reported in these proceedings are marked with filled triangles. Stars discovered earlier in the OGLE data are marked with asterisks and single star observed with {\it Kepler} is marked with filled circle.} 
\label{fig.pet068} 
\end{figure}

In Fig.~\ref{fig.pet068}, we show the Petersen diagram with all known stars of the discussed type. No specific structures are present within this group. 

A glimpse at the Petersen diagram (Fig.~\ref{fig.pet}) shows that the additional signal cannot correspond to any radial mode. The corresponding period is longer than the fundamental mode period. Moreover, it cannot correspond to stellar rotation or presence of a companion (it is too short). It seems that the most likely explanation is excitation of non-radial mode of gravity or of mixed character. This explanation faces many difficulties, however \citep[for a short discussion see section 5 in][]{068}. Recently, \cite{dziembowski} noted that additional periodicity could be attributed to radial fundamental mode if these objects are low-mass striped giants rather than genuine RRc stars. This explanation has its own problems as well, as discussed by \cite{dziembowski}. We have checked whether these stars are correctly identified as RRc pulsators. We investigated light curve shapes visually and quantitatively using the Fourier decomposition parameters \citep[see fig. 3 and 4 in][]{068}. They are typical for RRc stars. For a few stars we detect higher scatter of the phased light curve \citep[see fig.~3 in][]{068}. This is either due to lower mean brightness of these stars or due to period change which is common among RRc stars. Based on these considerations, identification of these stars as RR~Lyrae variables seems correct.

Two stars are interesting on their own and are discussed in more detail below.

\begin{figure}[!ht]
\centering
\includegraphics[width=0.8\textwidth]{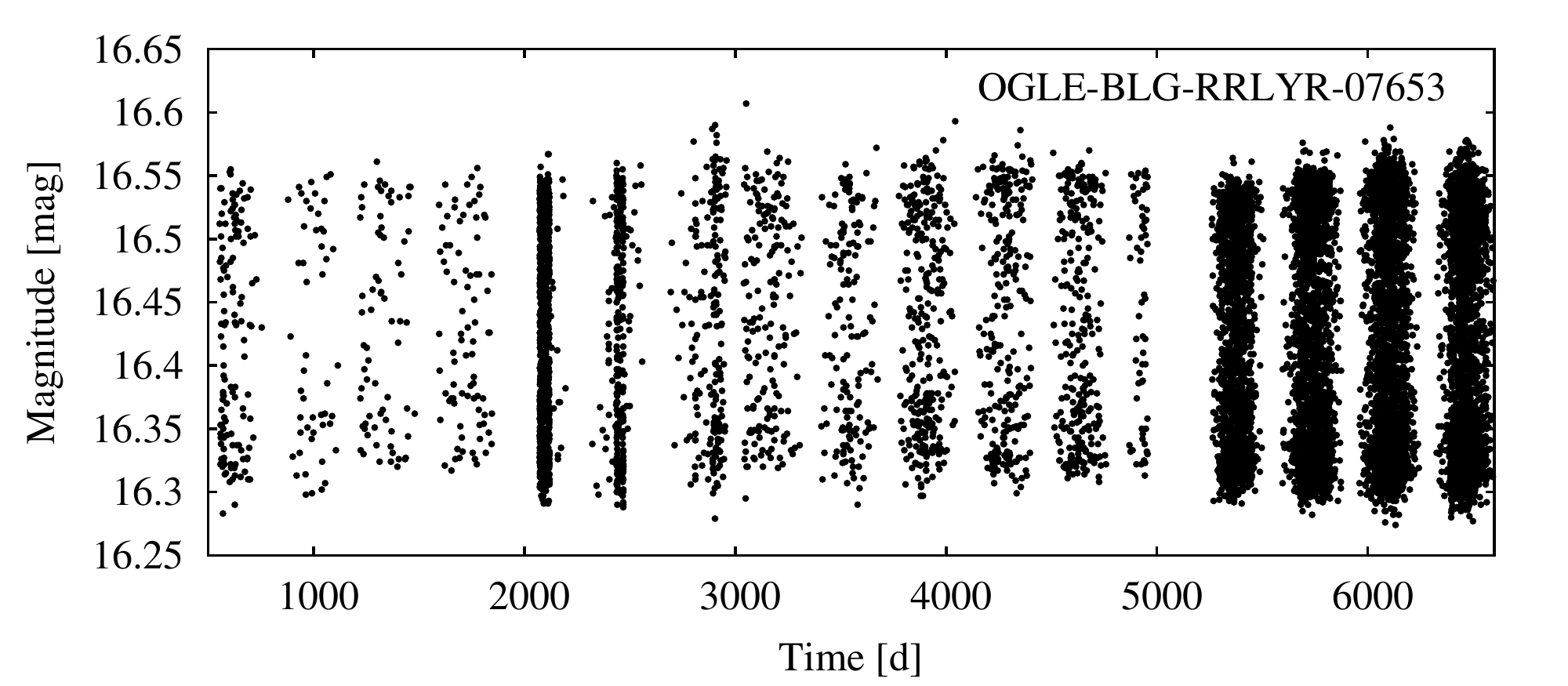}
\caption{Combined data from OGLE-IV, OGLE-III and OGLE-II for OGLE-BLG-RRLYR-07653. Total lengths of the data is $6042$\thinspace d. There are $10\,836$ data points.} 
\label{fig.07653}
\end{figure}

\begin{figure}[!ht]
\centering
\includegraphics[width=0.8\textwidth]{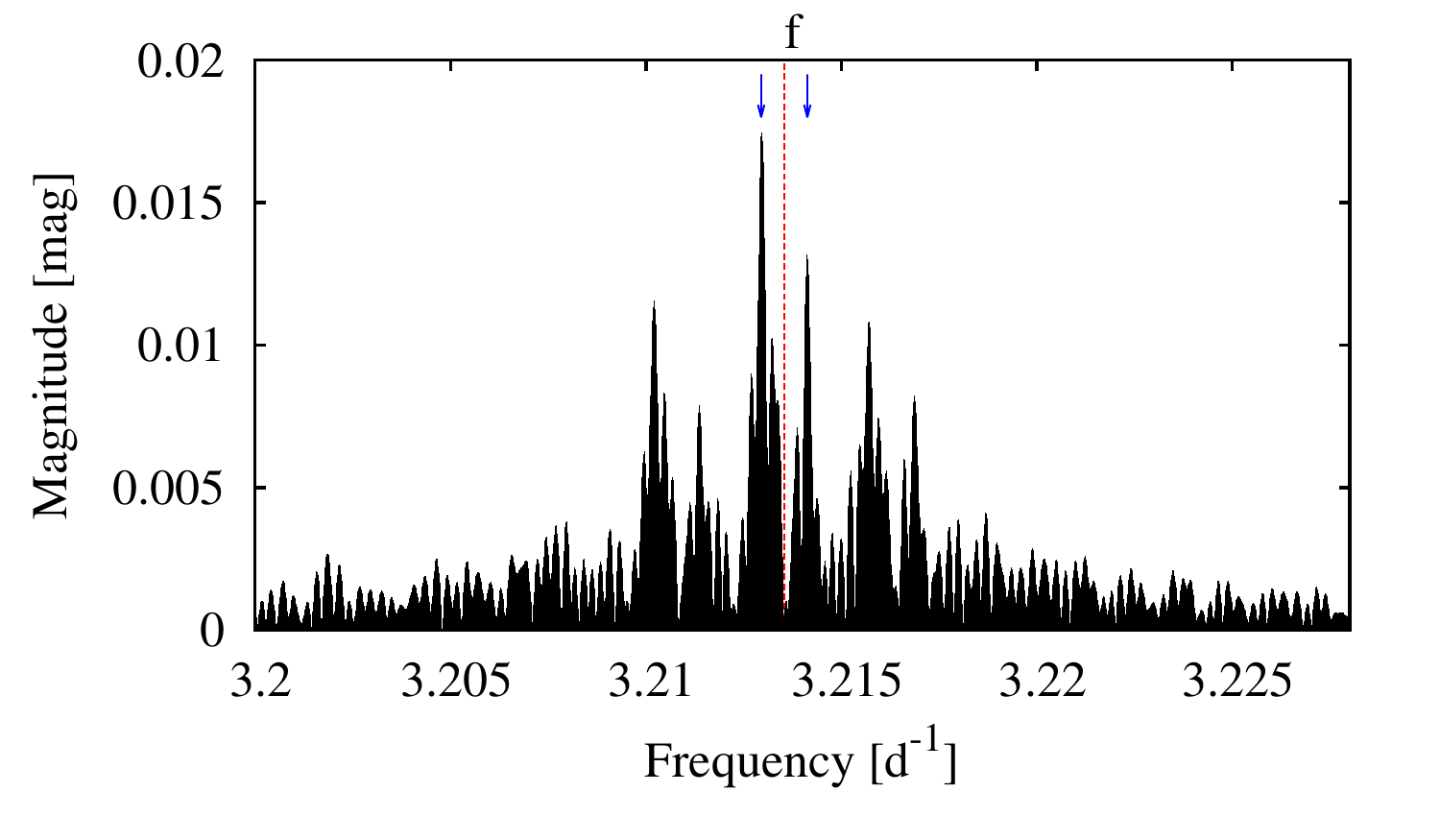}
\caption{Frequency spectrum of combined data for OGLE-BLG-RRLYR-07653 after prewhitening with the first overtone and its harmonics. Position of the first overtone is marked with $f$ and with dashed line. Side peaks are marked with arrows.} 
\label{fig.07653-fs}
\end{figure}

In the frequency spectrum of OGLE-BLG-RRLYR-07653, after prewhitening with the first overtone and its harmonics, we detected residual unresolved signal at the position of $f_{\rm 1O}$. We joined all available data for this star. Merged data are plotted in Fig.~\ref{fig.07653}. Increased length of the data set allowed us to detect triplet structure in the frequency spectrum. The side peaks at $f_{\rm 1O}$ are presented in Fig.~\ref{fig.07653-fs}. We interpret this structure as a signature of the Blazhko effect (long-period amplitude and/or phase modulation). Period of the Blazhko modulation is $1698\pm4$\thinspace d.

 In the frequency spectrum for OGLE-BLG-RRLYR-30601, we found more than one additional frequency. Additional long-period signal, $f_{\rm X}$, with $P_{\rm 1O}/P_{\rm X}\approx0.687$, is clearly detected, together with the combination frequency with the first overtone. There is another signal in the frequency spectrum at $f_{\rm Y}\approx 2.712544$\thinspace d$^{-1}$. Period ratio,  $P_{\rm 1O}/P_{\rm Y}\approx 0.7695$, is too high to fit the RRd sequence in the Petersen diagram. Combination frequency, $f_{\rm Y}+f_{\rm 1O}$, is also detected, which rules out the possible contamination. Signal at $f_{\rm Y}$ has higher amplitude than signal at $f_{\rm X}$. There are two other significant signals close to $f_{\rm 1O}$. Together with first overtone frequency they form unevenly spaced triplet. Separation between the low-frequency side peak and the first overtone is $0.07534$\thinspace d$^{-1}$, which corresponds to modulation period of $13.27$\thinspace d. For the high-frequency side peak the separation is $0.07173$\thinspace d$^{-1}$ (modulation period of $13.94$\thinspace d). This structure is not detected at harmonics of the first overtone. Light curve of this star, its Fourier decomposition parameters, are typical for RRc star.

\section{Summary}
We have analysed first overtone RR~Lyrae stars located in the Galactic bulge from the OGLE collection of variable stars. We focused on a search for additional periodicities in these stars. We have significantly increased the number of double-mode stars with $P_{\rm X}/P_{\rm 1O}\approx 0.61$. More than 260 stars were detected in the OGLE data \citep{061oiii,061oiv}. We have also discovered entirely new group of double-periodic stars, with dominant variation associated with the radial first overtone and additional periodicity of longer period \citep{068}. In the Petersen diagram, these stars form a horizontal sequence, with period ratios, $P_{\rm 1O}/P_{\rm X}$, clustering around $0.686$. So far we know 20 stars of this type. These stars still lack a satisfactory explanation.
 
\section*{Acknowledgements}
This research is supported by the Polish National Science Center through grant DEC-2012/05/B/ST9/03932.

\bibliographystyle{ptapap}
\bibliography{netzel}

\begin{thebibliography}{9}
\providecommand{\natexlab}[1]{#1}
\providecommand{\url}[1]{\texttt{#1}}
\providecommand{\urlprefix}{URL }
\providecommand{\eprint}[2][]{\url{#2}}

\bibitem[{{Dziembowski}(2015)}]{dziembowski}
{Dziembowski}, W.~A., \emph{{Nonradial Oscillations in Classical Pulsating
  Stars. Predictions and Discoveries}}, \emph{ArXiv e-prints}  (2015),
  \eprint{1512.03708}

\bibitem[{{Jurcsik} et~al.(2015)}]{jurcsik_M3}
{Jurcsik}, J., et~al., \emph{{Overtone and Multi-mode RR Lyrae Stars in the
  Globular Cluster M3}}, \emph{\apjs} \textbf{219}, 25 (2015),
  \eprint{1504.06215}

\bibitem[{{Moskalik} et~al.(2015)}]{pamsm15}
{Moskalik}, P., et~al., \emph{{Kepler photometry of RRc stars: peculiar
  double-mode pulsations and period doubling}}, \emph{\mnras} \textbf{447},
  2348 (2015), \eprint{1412.2272}

\bibitem[{{Netzel} et~al.(2015{\natexlab{a}}){Netzel}, {Smolec}, \&
  {Dziembowski}}]{068}
{Netzel}, H., {Smolec}, R., {Dziembowski}, W., \emph{{Discovery of a new group
  of double-periodic RR Lyrae stars in the OGLE-IV photometry}}, \emph{\mnras}
  \textbf{451}, L25 (2015{\natexlab{a}}), \eprint{1504.05765}

\bibitem[{{Netzel} et~al.(2015{\natexlab{b}}){Netzel}, {Smolec}, \&
  {Moskalik}}]{061oiii}
{Netzel}, H., {Smolec}, R., {Moskalik}, P., \emph{{Double-mode
  radial-non-radial RR Lyrae stars in the OGLE photometry of the Galactic
  bulge}}, \emph{\mnras} \textbf{447}, 1173 (2015{\natexlab{b}}),
  \eprint{1411.3155}

\bibitem[{{Netzel} et~al.(2015{\natexlab{c}}){Netzel}, {Smolec}, \&
  {Moskalik}}]{061oiv}
{Netzel}, H., {Smolec}, R., {Moskalik}, P., \emph{{Double-mode
  radial-non-radial RR Lyrae stars. OGLE-IV photometry of two high cadence
  fields in the Galactic bulge}}, \emph{\mnras} \textbf{453}, 2022
  (2015{\natexlab{c}}), \eprint{1507.08414}

\bibitem[{{Soszy{\'n}ski} et~al.(2011)}]{ogle-iii-bulge}
{Soszy{\'n}ski}, I., et~al., \emph{{The Optical Gravitational Lensing
  Experiment. The OGLE-III Catalog of Variable Stars. XI. RR Lyrae Stars in the
  Galactic Bulge}}, \emph{\actaa} \textbf{61}, 1 (2011), \eprint{1105.6126}

\bibitem[{{Soszy{\'n}ski} et~al.(2014)}]{ogle-iv-bulge}
{Soszy{\'n}ski}, I., et~al., \emph{{Over 38000 RR Lyrae Stars in the OGLE
  Galactic Bulge Fields}}, \emph{\actaa} \textbf{64}, 177 (2014),
  \eprint{1410.1542}

\bibitem[{{Udalski} et~al.(2015){Udalski}, {Szyma{\'n}ski}, \&
  {Szyma{\'n}ski}}]{ogle-iv}
{Udalski}, A., {Szyma{\'n}ski}, M.~K., {Szyma{\'n}ski}, G., \emph{{OGLE-IV:
  Fourth Phase of the Optical Gravitational Lensing Experiment}}, \emph{\actaa}
  \textbf{65}, 1 (2015), \eprint{1504.05966}

\end{thebibliography}

\end{document}